\documentstyle[prl,aps,twocolumn,epsfig]{revtex}

\begin{document}

\twocolumn[\hsize\textwidth\columnwidth\hsize\csname @twocolumnfalse\endcsname

\title{Quantum resonances and decay of a chaotic fractal
repeller observed using microwaves}

\author{Wentao Lu, M.Rose, K.Pance and S.Sridhar$^{a}$}
\address{Physics Department, 360 Huntington Avenue, Boston, MA 02115.}

\date{\today}

\maketitle

\begin{abstract}
  The quantum resonances of classically chaotic n-disk geometries were
  studied experimentally utilizing thin 2-D microwave geometries. 
  The experiments yield the frequencies and widths of low-lying 
  resonances, which are compared with semiclassical calculations. 
  The longtime or small energy behavior of the wave-vector 
  auto-correlation gives information about the quantum decay rate, 
  which is in good agreement with that obtained from classical
  scattering theory. The intermediate energy behavior shows non-universal
  oscillations determined by periodic orbits.
\end{abstract}


\vskip1.5pc]

The n-disk scattering problem is one whose quantum-classical correspondence
has received extensive theoretical attention\cite{Cvitanovic88,Gaspard89},
because it is a paradigm of an open quantum chaos system, much as the
pendulum is for integrable systems. Furthermore it is relevant to physical
situations in diverse fields, such as the crossroads geometry for electron
devices \cite{Jalabert90}, unimolecular chemical reactions \cite{Gaspard89}
and electromagnetic and acoustic scattering \cite{Decanini98}.

The classical scattering function of n-disks on a plane is
non-differentiable and forms a Cantor set. A central property is the
exponential decay of an initial distribution of classical particles, hence
the name (fractal) repeller. For closed quantum chaotic billiards,
experimental and theoretical work is available on eigenvalues (which are
purely real) and eigenfunctions, and the many features of universality and
non-universal behavior are known \cite{Bohigas90}. In contrast, for open
quantum systems, the eigenvalues are intrinsically complex and their
universal behavior is a question of great interest \cite{John91}. Despite
extensive theoretical treatment there have been almost no real experiments
on the n-disk geometry which exemplify this unique problem in quantum chaos.

In this paper we present a microwave realization of the chaotic n-disk
problem. Experiments were carried out for $n=1,2,3,4,6$, as well as for
large $n=20$, the latter corresponding to the random Lorentz scatterer. In
this paper we focus on the case $n=4$. The experiments yield the frequencies
and the widths of the low lying quantum resonances of the four-disk
repeller. We have also carried out semiclassical calculations of the
resonances, which are shown to reproduce the resonances reasonably well. Our
experiments enable us to explore the role of symmetry in a unique way by
studying different irreducible representations. The experimental data are
used to display the signatures of the classical chaos in the transmission
spectra, through measures such as the spectral (wave-vector $k$)
auto-correlation function. The small $k$ (long time) behavior of this
quantity provides a measure of the quantum escape rate, and is shown to be
in good agreement with the corresponding classical escape rate. For large $k$
(short times), the contribution of periodic orbits is observed as
non-universal oscillations of the auto-correlation.

The experiments are carried out in thin microwave structures consisting of
two highly conducting $Cu$ plates spaced $d\,\sim 6\,mm$ and about $55\times
55\,cm$ in area. Discs and bars also made of $Cu$ and of thickness $d$ are
placed between the plates and in contact with them. In order to simulate an
infinite system microwave absorber material was sandwiched between the
plates at the edges. Microwaves were coupled in and out using loops
terminating coaxial lines which were inserted in the vicinity of the
scatterers. All measurements were carried out using an HP8510B vector
network analyzer which measured the complex transmission ($S_{21}$) and
reflection ($S_{11}$) S-parameters of the coax + scatterer system. It is
crucial to ensure that there is no spurious background scattering due to the
finite size of the system. This was verified carefully as well as that the
effects of the coupling probes were minimal and did not affect the results.

In this essentially 2-D geometry, Maxwell's equation for the experimental
system is identical with the Schr\"odinger time-independent wave equation 
$(\nabla ^{2}+k^{2})\Psi =0$ with $\Psi =E_{z}$ the $z$-component of the
microwave electric field. This correspondence is exact for all frequencies 
$f_{c}<c/2d=25GHz$.(Note that $k=2\pi f/c$, where $c$ is the speed of light).
It is this mapping which enables us to study the quantum properties of the
n-disk system. For all metallic objects in the 2-D space between the plates,
Dirichlet boundary conditions apply inside the metal. Thus for the 4-disk
geometry, $\Psi =0$ inside the 4 disks.

The transmission function $S_{21}(f)$ which we measure is the response of
the system to a delta-function excitation at point $\vec{r}_{1}$ probed at a
different point $\vec{r}_{2}$, and is determined by the wavefunction $\Psi $
at the probe locations $\vec{r}_{1}$and $\vec{r}_{2}$. In our experiments
the coax lines act as tunneling point contacts, and hence it can be shown 
\cite{Lu98} that $S_{21}(f)=A(f)G(\vec{r}_{1},\vec{r}_{2},f)$ is just the
two-point Green's function $G(\vec{r}_{1},\vec{r}_{2},f)$, scaled by a
slowly varying function $A(f)$ of frequency $f$ which represents the
impedance characteristics of the coax lines and probes. Because we ensure
that the coupling to the leads is very weak, any shifts due to the leads are
negligible ($<10^{-4}$ of the resonance frequencies and widths)\cite{Stein95}.

Note that the measurements of $|S_{21}(f)|^{2}$ are equivalent to a
two-probe measurement of the conductance \cite{Jalabert90}, since the
conductance $g\propto \sum |t_{nm}|^{2}$, the sum of the transmission
probabilities $t_{nm}$ for all input/output channels $(n,m)$. In the present
experiment, the input/output leads are $\delta $-function point contacts,
and hence $t_{11}=S_{21}$and $g\propto |S_{21}(f)|^{2}$. Thus our
experiments enable us to make comparison with theories \cite
{Jalabert90,Alhassid98,Jensen96,Delos96} originally developed for electronic
micro-structures \cite{Marcus92,Chang94}.

\begin{figure}[h]
\epsfig{width=.9 \linewidth,file=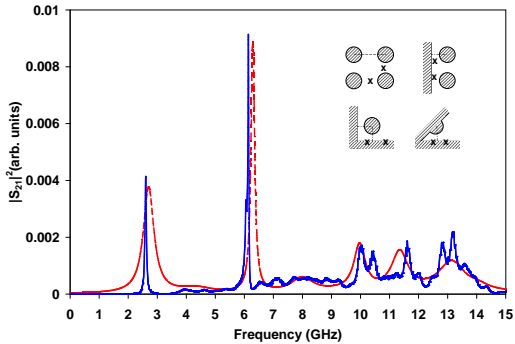,angle=0}
\vspace{0.0truecm}
\vskip-.4cm
\caption{Transmission function $|S_{21}|^2$ for a 4-disk system with 
$R=8cm$ and $a=2cm$.
The dashed line is the semi-classical calculation.
(Inset) the experimental configuration and the various reduced representations of the 4-disk
geometries studied in this work. The separation distance $R$ is shown as a dashed line.
The x's mark typical locations of the coupling probes. }
\label{fig1}
\end{figure}

The transmission function $|S_{21}|^{2}$ for a square 4-disk system with
disk radius $a=2cm$ and separation $R=8\,cm$ is shown in Fig.\ref{fig1}. The
data shown is the sum of traces taken for 3 locations of the probes, in
order to avoid accidental zeros of the wavefunction at the probe locations.
Several resonances are clearly seen. The resonance frequencies $f$ and
widths $\Delta f$ directly yield the complex wave-vector quantum eigenvalues
of the 4-disk repeller. In our previous experiments on closed cavities such
as Sinai billiard geometries\cite{Sridhar91}, the widths of the
eigen-resonances were due to dissipation in the metal walls. Here however
the resonance widths are due to decay of the wavefunction into infinite
space. These experiments thus enable us to explore quantum decay in open
systems. It must be emphasized that the dissipation in the walls is entirely
negligible in the present experiments.

\vskip-.4cm

The experimental approach was validated by measurements (not shown) on an
(integrable) 2-disk system, for which we have measured both in the full
space with two actual disks ($A_{1}$ and $A_{2}$ representations) and in the
half-space using a reflecting mirror ($A_{2}$ representation). Here the
spectrum consists of a 1-parameter family of resonances due to a single
periodic orbit. The experimental data are in very good agreement with
semiclassical calculations (for preliminary measurements, see \cite
{Kudrolli95}).

For the chaotic 4-disk case, we have also carried out semiclassical
calculations using methods in the literature \cite
{Cvitanovic88,Gaspard89,Gaspard92,Gaspard94,Gaspard98,Cvitanovic98,Cvitanovic89}%
. In the semiclassical theory, the resonances of the system are given by the
poles of the Ruelle $\zeta $ function with $j$ running from $0$ to $\infty $%
\cite{Gaspard94} 
\begin{equation}
\zeta _{(1/2)+j}(-ik)=\prod_{p}\left[ 1-(-1)^{L_{p}}e^{ikl_{p}}/
\Lambda _{p}^{(1/2)+j})\right] ^{-1}
\end{equation}
where $k$ is the wave vector, $l_{p}$ is the length of the periodic 
orbit $p$
, $L_{p}$ the number of collisions of the periodic orbit with the disks, and 
$\Lambda _{p}$ is the eigenvalue of the stability matrix.

For the 4-disk system with $C_{4v}$ symmetry, the semi-classical
calculations using the cycle expansions \cite{Cvitanovic88} were carried out
up to period 3 including 14 periodic orbits. Because of the $C_{4v}$
symmetry, there are five representations of the Ruelle $\zeta $ function\cite
{Cvitanovic89}, $A_{1}$, $A_{2}$, $B_{1}$, $B_{2}$, $E$, where the last one
is two-dimensional. The poles of the first Ruelle $\zeta $ function with 
$j=0 $ are calculated, since they contribute to the resonances with the
longest lifetimes. The energy of the particle is $E=k^{2}$(here we can take 
$m=1/2,\hbar =1$), and is also complex.

In Fig.\ref{fig1}, we show the results of the semi-classical calculations,
as a superposition of Lorentzians 
$S|_{cal}^{2}(k)=\sum_{i}c_{i}\gamma _{i}/((k-s_{i})^{2}+\gamma _{i}^{2})$,
where $s_{i}$ and $\gamma _{i}$ are the calculated real and imaginary parts
of resonances, respectively. The coupling constants $c_{i}$ were chosen to
fit the data - as explained above they are specific to the probe locations.
Good agreement is found for the resonance frequencies of the sharp
resonances. The sharp resonances with small imaginary parts and hence high 
$Q$ are easily recognized while the resonances with large imaginary parts
are not easy to distinguish, although all resonances contribute to the
transmission function in Fig.1.

The approximate semiclassical theory used provides a fair prediction of the
resonance frequencies even for the low-lying resonances. The accuracy is
within $5\%$. But the agreement for the widths is not as good as that for
the resonance frequencies. The lower the frequency, the greater the
discrepancy between the calculated and experimental widths. The discrepancy
may improve if more orbits of higher period are included. Another source of
the discrepancy is intrinsic in the semi-classical theory because of the
large correction of the stationary phase approximation\cite{Wirzba92}.
Comparison with exact quantum calculations\cite{Gaspard94,Decanini98} would
be desirable.

One of the noteworthy features of the experiments is the ability to vary
geometry. We exploited this by exploring the role of the symmetry in the
4-disk geometry. We performed the experiment in four different setups (see
inset to Fig.\ref{fig1}). These correspond to four different ways of probing
the phase space: the full space in which all five representations are
included, half $(A_{2},B_{2},E)$, one fourth $(B_{1},B_{2})$, and one eighth
of the space ($B_{2})$. Approximately 17 configurations were studied and
analyzed in details. The systematic trends are consistent with expectations.
Here we discuss some of the principal features of the results, further
extensive details will be published in a longer publication \cite{Lu98}.

\begin{figure}[h]
\epsfig{width=.9 \linewidth,file=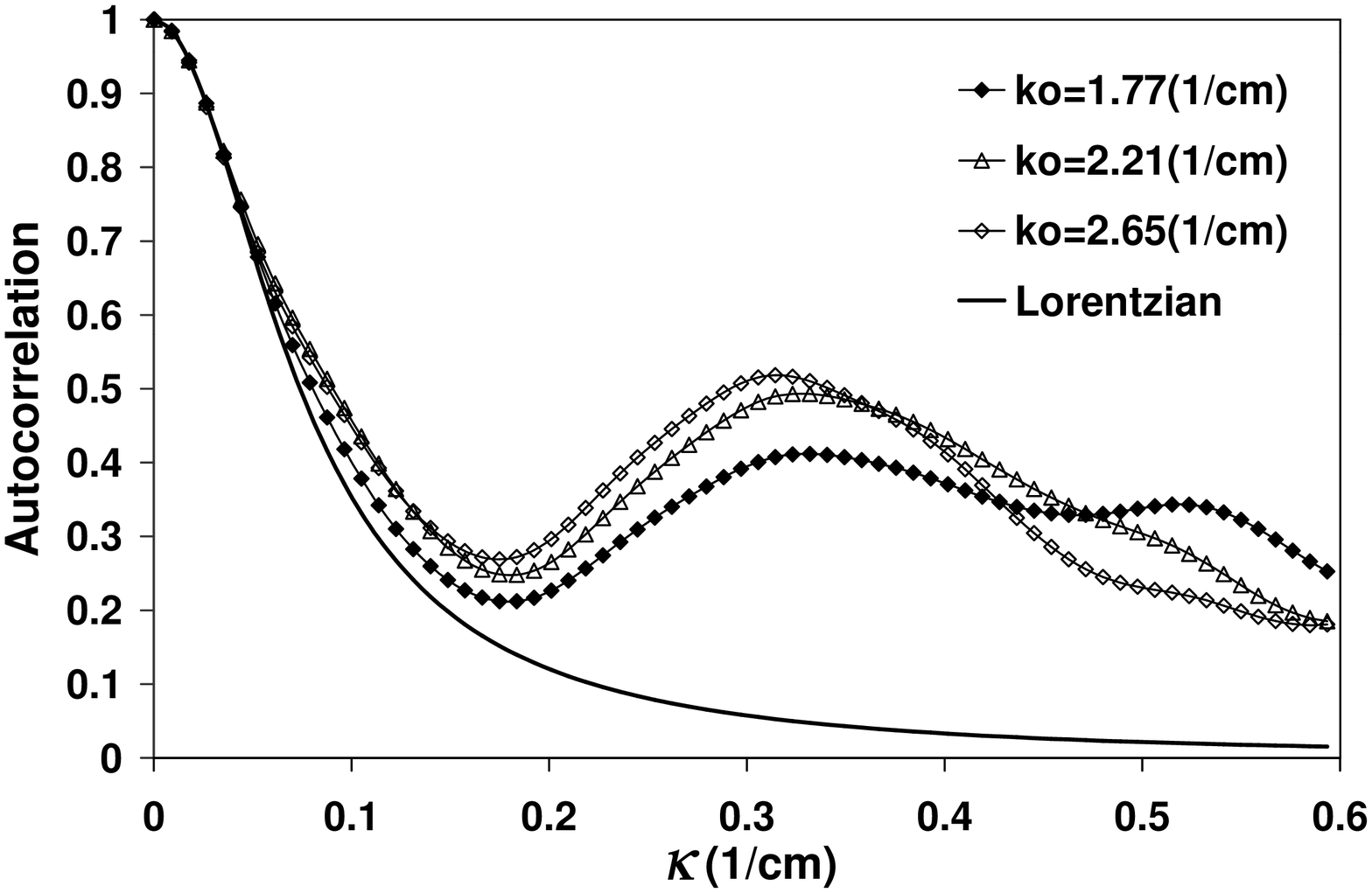,angle=0}
\vspace{0.0truecm}
\vskip-.4cm
\caption{Wave-vector autocorrelation $C(\kappa )$ of the 4-disk system 
  with $R=20cm$ and $a=5cm$. Data are shown for 1/8th configuration of 
  the 4-disk geometry corresponding to the $B_2$ representation.
  The correlation is calculated with interval $\Delta k=2 cm^{-1}$. 
  The different sets represent different values of the central 
  wave-vector $k_0$. The bold line is a Lorentzian with 
  $\gamma_{qm}=0.075 cm^{-1}$.}
\label{fig2}
\end{figure}

We now turn to another analysis of the data in terms of the spectral
auto-correlation function, which was calculated as 
$C(\kappa)=<|S_{21}(k-(\kappa /2))|^{2}|S_{21}(k+(\kappa /2))|^{2}>_{k}$. The average
is carried out over a band of wave vector centered at certain value $k_{0}$
and of width $\Delta k$. The window function\cite{Lai92} chosen was 
$f(x)=(1-|x|/\sqrt{6})/\sqrt{6}$ for $|x|<\sqrt{6}$ and $f(x)=0$ for $|x|\ge 
\sqrt{6}$ with $x=(k-k_{0})/\Delta k$. The trace $S_{21}$ used is the
average of several traces collected at different probe locations with the
same geometrical configuration of the disks to avoid missing resonances due
to the accidental coincidence of either probe with a node of the
wavefunction. The autocorrelation is the average of that for several $k_{0}$. 
A plot of the autocorrelation for the 1/8th configuration of the 4-disk
geometry is shown in Fig.\ref{fig2}.

\vskip-.4cm

Since $|S_{21}(k)|^{2}=\sum_{i}c_{i}\gamma _{i}/((k-s_{i})^{2}+\gamma
_{i}^{2})$, we have\cite{Eckhardt93} $C(\kappa )=\pi
\sum_{i,j}c_{i}c_{j}(\gamma _{i}+\gamma _{j})/((\kappa
-(s_{i}-s_{j}))^{2}+(\gamma _{i}+\gamma _{j})^{2}).$ In the case that there
are no overlapping resonances, $|s_{i}-s_{j}|>>(\gamma _{i}+\gamma _{j})$,
the small $\kappa $ behavior of the autocorrelation is $C(\kappa )\approx
\pi \sum_{i}2c_{i}^{2}\gamma _{i}/(\kappa ^{2}+4\gamma _{i}^{2}).$

According to semiclassical theory, the above sum can be replaced by a single
Lorentzian \cite{Blumel88,Jalabert90,Lewenkopf91,Lai92} 
\begin{equation}
C(\kappa )=C(0){\frac{1}{1+(\kappa /\gamma )^{2}}}.
\end{equation}
In semi-classical theory, $\gamma =\gamma _{0}$, the classical escape rate
with the velocity scaled to 1. Thus one can interpret the width of the
autocorrelation as an average lifetime of resonances
\cite{Ericsson60,Brink63}. The above equation was used to fit the spectral autocorrelation for small $%
\kappa $ and thus obtain the value of the experimental escape rate $\gamma
_{qm}$, as shown in Fig.\ref{fig2}.

In the classical scattering theory, the classical escape rate $\gamma _{0}$
is the simple pole of the classical Ruelle $\zeta $ function 
\cite{Gaspard92}. We use the results of ref.\cite{Gaspard92} in Fig.3 for comparison with
the experimental data. Asymptotically, when $R$ is large, $\gamma
_{0}\approx \ln (2\sqrt{2}R/a)/(\sqrt{2}R)$. Another relevant quantity is
the abscissa of absolute convergence $s_{c}$ for Eq. (1) which can also be
estimated from the Ruelle $\zeta $ function with the classical cycle weights 
$t_{p}$ replaced by the corresponding semiclassical ones. $s_{c}$ serves as
a lower bound of the escape rate\cite{Eckhardt95}, and is also shown in
Fig.3. (Note that this is negative for $R/a<4.5$).

\vskip-.4cm
\begin{figure}[h]
\epsfig{width=.9 \linewidth,file=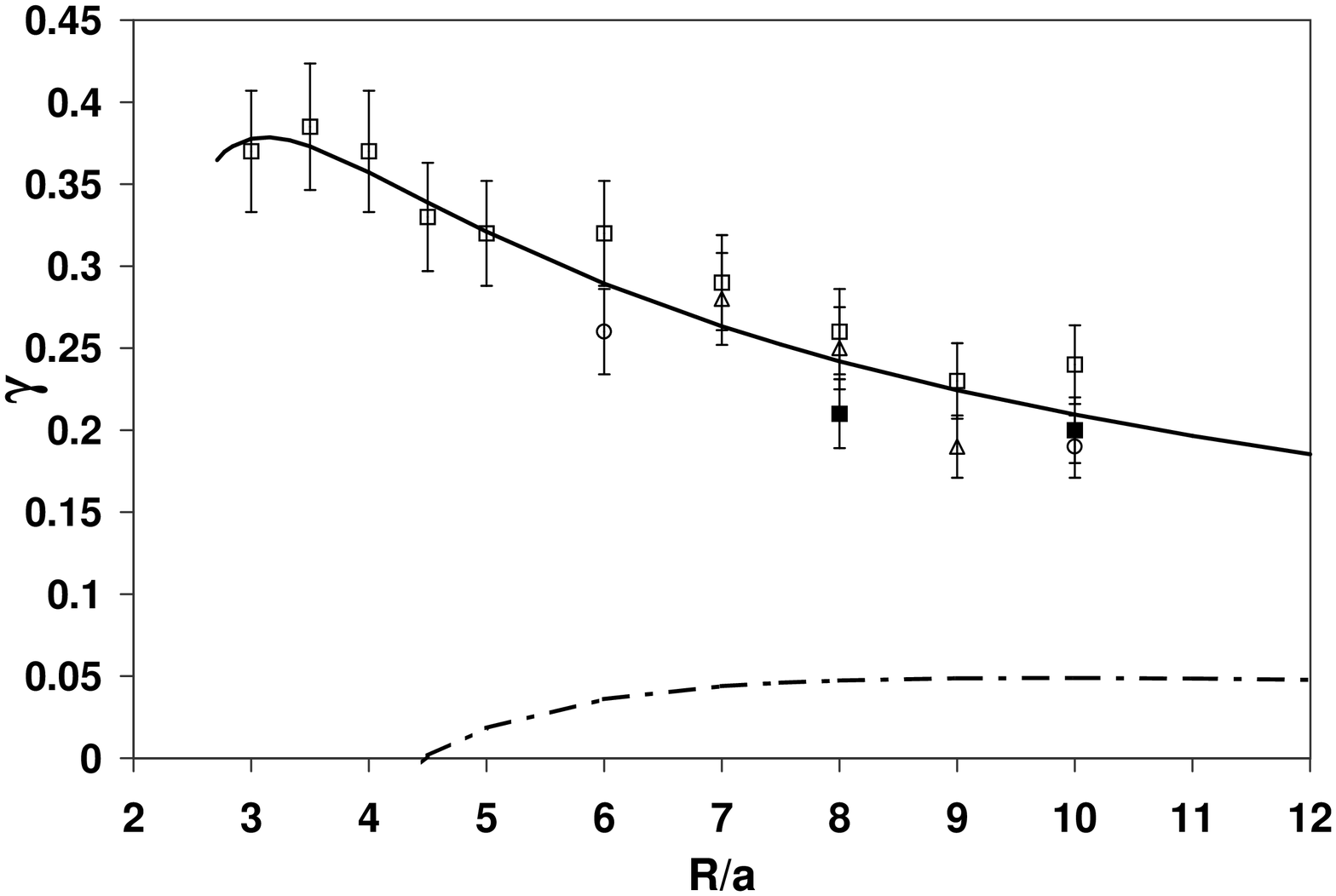}
\vspace{0.0truecm}
\vskip-.4cm
\caption{Experimental escape rate $\gamma _{qm}$ scaled to radius $a = 1$ versus $R/a$. Data are shown for 
different reduced configurations of the 4-disk geometry : 1/8th space(open squares),
$1/2$ space (open circles), $1/4$ space (filled squares), full space (triangles).
The classical escape rate (solid line) is calculated from the 
first three periodic orbits in the fundamental domain. The abcissa of convergence $s_c$ of Eq.(1) is 
shown as a dot-dashed line, and represents a lower bound on the quantum escape rate.}
\label{fig3}
\end{figure}

Good agreement of the escape rate is obtained between that of the classical
theory $\gamma _{0}$ and that from the experimental data $\gamma _{qm}$.
This is shown in Fig.3, where we compare the experimental escape rates $%
\gamma _{qm}$ with the classical escape rate $\gamma _{0}$ for several
values of $R/a$. Note that in Fig.3, data are included for 17 configurations
of the different reduced ($1/8$, $1/4$, $1/2$ and full space)
representations of the 4-disk geometry shown in Fig.1. The radius of the
disks used was $a=5cm$ for the $1/8$ space, and $a=2cm$ for the others. The
data for $\gamma _{qm}$ are scaled to radius $a=1$. In Fig.3 we have shown
experimentally that the small $\kappa $ behavior of the spectral
auto-correlation has a universal behavior in that it is independent of the
details of the geometry. The good agreement in Fig.3 between the measured
escape rate $\gamma _{qm}$ and the classical escape rate $\gamma _{0}$ shows
that some quantum properties are well predicted by semiclassical theory.

For intermediate $\kappa $, the semiclassical theory Eq.(2) fails because of
the presence of the periodic orbits, which lead to non-universal behavior.
In the case of just one periodic orbit, $C(\kappa )\propto
\sum_{n=0}^{\infty }{2\gamma /[}(\kappa -n\Delta s)^{2}+4\gamma ^{2}]$. For
example for the 2-disk problem if just the $A_{2}$ representation is present 
\cite{Kudrolli95}, $\Delta s=2\pi /(R-2a)$, $2\gamma =\ln \Lambda /(R-2a)$
is the width of the resonances with the eigenvalue of the instability matrix 
$\Lambda =[R-a+\sqrt{R(R-2a)}]/a$. $2\gamma $ is also the classical escape
rate $\gamma _{0}$ of the system. Thus the autocorrelation oscillates
exactly with period $\Delta s$. Excellent agreement is found between
experiment and theory for this 2-disk case.

In the full space of the 4-disk system, the average length of the periodic
orbits per period can be estimated as the average length of the eight
periodic orbits, $12$, $23$, $34$, $41$, $1234$, $1432$, $13$, $24$, where 
$1,2,3,4$ are the labels of the four disks\cite{Cvitanovic89}. The mean
separation between the resonances is approximately given by 
$\overline{\Delta s}=2\pi /[(2+1/\sqrt{2})R-(3+\sqrt{2})a]$. For the 4-disk system with
one-eighth of the phase space, the mean separation is 
$\overline{\Delta s}
=6\pi /[(3+\sqrt{2})R-2(2+\sqrt{2})a]$. The autocorrelation will oscillate
with approximate period $\overline{\Delta s}$, which indicates the deviation
from the semiclassical theory because of the presence of the periodic orbits
in the system. Thus the short time behavior is system specific. The value of 
$\overline{\Delta s}=0.35\,cm^{-1}$ is in good agreement with the scale of
oscillations in Fig.2. Future work will focus on the interplay between the
universal chaotic features and the non-universal PO contributions.

It is interesting to examine the variation of the results with increasing $n$.
 For $n=\infty $ one obtains the Lorentz scatterer. For $n=20$, we find
that $\gamma _{fit}\sim 0.05$, indicating that as $n$ increases the system
approaches a closed system and the escape rate becomes very small.

To our knowledge this work represents the first determination of the
wave-vector autocorrelation for an experimental system. Our results thus
nicely complement measurements in semiconductor microstructures \cite
{Marcus92,Jalabert90,Chang94}, where however the wave-vector correlation is
difficult to extract but instead the magnetic field correlations of
conductivity $g(\Delta B)$ are analyzed. As noted previously, the present
experiment exactly corresponds to an ideal non-interacting electron in a
quantum dot, with tunneling contacts. Thus Fig.1 can be viewed as equivalent
to conductivity fluctuations in a quantum dot.

The present work has provided the first experimental realization of the
n-disk open billiards problem. {\em The experiments clearly demonstrate the
quantum-classical correspondence, in that the} {\em quantum properties can
be obtained from classical quantities and vice-versa.} One of the powerful
features of the present experiments is the (almost unlimited) ability to
vary geometry, as we have already demonstrated by studying numerous
configurations. Hence there is enormous potential for addressing a variety
of issues in the quantum-classical correspondence problem. In addition open
billiards are also a paradigm for dissipative quantum systems, a problem of
wide importance. Besides their relevance to atomic and chemical physics, the
present results also have applicability to situations where wave mechanics
plays a role, such as in electromagnetism and acoustics.

This work was supported by NSF-PHY-9722681.We thank A. Kudrolli, V. Kidambi,
J. V. Jose, N. Whelan and L.Viola for useful discussions.

$^{a}$ electronic address : srinivas@neu.edu.

\end{document}